\documentstyle[11pt,amsfonts]{article}

\def\hybrid{\topmargin 0pt      \oddsidemargin 0pt
	\headheight 0pt \headsep 0pt
	\textheight 9in         
	\textwidth 6.25in       
	\marginparwidth .875in
	\parskip 5pt plus 1pt   \jot = 1.5ex}

\catcode`\@=11
\def\marginnote#1{}
\newcount\hour
\newcount\minute
\newtoks\amorpm
\hour=\time\divide\hour by60
\minute=\time{\multiply\hour by60 \global\advance\minute by-\hour}
\edef\standardtime{{\ifnum\hour<12 \global\amorpm={am}%
	\else\global\amorpm={pm}\advance\hour by-12 \fi
	\ifnum\hour=0 \hour=12 \fi
	\number\hour:\ifnum\minute<10 0\fi\number\minute\the\amorpm}}
\edef\militarytime{\number\hour:\ifnum\minute<10 0\fi\number\minute}

\def\draftlabel#1{{\@bsphack\if@filesw {\let\thepage\relax
  \xdef\@gtempa{\write\@auxout{\string
   \newlabel{#1}{{\@currentlabel}{\thepage}}}}}\@gtempa
  \if@nobreak \ifvmode\nobreak\fi\fi\fi\@esphack}
	\gdef\@eqnlabel{#1}}
\def\@eqnlabel{}
\def\@vacuum{}
\def\draftmarginnote#1{\marginpar{\raggedright\scriptsize\tt#1}}

\def\draft{\oddsidemargin -.5truein
	\def\@oddfoot{\sl preliminary draft \hfil
	\rm\thepage\hfil\sl\today\quad\militarytime}
	\let\@evenfoot\@oddfoot \overfullrule 3pt
	\let\label=\draftlabel
	\let\marginnote=\draftmarginnote
  \def\@eqnnum{(\theequation)\rlap{\kern\marginparsep\tt\@eqnlabel}%
\global\let\@eqnlabel\@vacuum} }

\def\numberbysection{\@addtoreset{equation}{section}
	\def\theequation{\thesection.\arabic{equation}}}

\def\underline#1{\relax\ifmmode\@@underline#1\else
	$\@@underline{\hbox{#1}}$\relax\fi}

\def\titlepage{\@restonecolfalse\if@twocolumn\@restonecoltrue\onecolumn 
   \else \newpage \fi \thispagestyle{empty}\c@page\z@
	\def\thefootnote{\fnsymbol{footnote}}
	\setcounter{footnote}{1}}	

\def\endtitlepage{\if@restonecol\twocolumn \else \fi
	\def\thefootnote{\arabic{footnote}}
	\setcounter{footnote}{0}} 
\catcode`@=12
\relax
\def\ie{\hbox{\it i.e.}\/}    
    
\def\etal{\hbox{\it et al.}\/}

\def\beq{\begin{equation}}
\def\eeq{\end{equation}}
\def\bea{\begin{eqnarray}}
\def\eea{\end{eqnarray}}
\def\bar{\overline}

\def\pa{\partial}
\def\d{{\cal d}}

\def\n{ {\psi}}

\def\m{\mu}
\def\n{\nu}
\def\r{\rho}
\def\s{\sigma}

\def\e{\epsilon}

\def\a{\alpha}
\def\b{\beta}

\def\d{\delta}
\def\l{\lambda}
\def\demi{{1\over 2}}

\def\NN{N_{eq}}

\def\nn{\nonumber}

\def\udl{\underline}
\def\trep{\udl{$d_T$}\/}
\def\ttrep{\udl{$d_{T^2}$}\/}
\def\ttrepm{\udl{$d_{T_m}$}\/}
\def\frep{\udl{$d_F$}\/}
\newcommand{\rep}[1]{\udl{$#1$}\/}

\def\au{$su_2$} 

\def\at{$su_4$} 
\def\aq{$su_5$}
\def\ac{$su_6$}
\def\as{$su_7$}
\def\ase{$su_8$}
 
\def\bd{$so_5$} 
\def\bt{$so_7$}

\def\bc{$so_{11}$}

\def\bse{$so_{15}$}
\def\cd{$sp_4$} 
\def\ct{$sp_6$}
\def\cq{$sp_8$}

\def\dt{$so_6$}
\def\dq{$so_8$}
\def\dc{$so_{10}$}
\def\ds{$so_{12}$}
\def\dse{$so_{14}$}
\def\dh{$so_{16}$}

\newcommand{\deb}[1]{\begin{minipage}{#1em} }
\newcommand{\fin}{\end{minipage}}

\def\sde{\hbox{self-duality equation}}
\def\sdes{\hbox{self-duality equations}}

\def\st{\hbox{space-time}}

\def\repr{\hbox{representation}}
\def\irr{\hbox{irreducible}}
\def\reprs{\hbox{representations}}

\def\sod{\hbox{$so_D$}}
\def\Sod{\hbox{$SO_D$}}

\def\sg{\hbox{sub-group}}

\def\sa{\hbox{sub-algebra}}
\def\sas{\hbox{sub-algebras}}

\relax
\numberbysection
\hybrid
\begin{document}
\begin{titlepage}
\begin{center}
\hfill {\small PAR--LPTHE 98--01} \\
\hfill {\small CERN--TH 98--04} \\
 [4em]
{\large\bf ON GENERALIZED SELF-DUALITY EQUATIONS TOWARDS SUPERSYMMETRIC
QUANTUM FIELD THEORIES OF FORMS}\\[4em]
 {\bf Laurent Baulieu$^{\dag \ddag}$ and C\'eline Laroche$^{\dag}$} \\
(baulieu, laroche@lpthe.jussieu.fr)\\ [.2in]

{\footnote{\noindent URA CNRS 280, Boite 126, Tour 16, 1$^{\it er}$
\'etage, 4 place Jussieu, F-75252 Paris Cedex 05, France.}LPTHE, Paris,
France.}\\
{Universit\'es Pierre et Marie Curie (Paris 6) et Denis Diderot (Paris
7)}\\ and\\ 
{\footnote{TH-Division, CERN, 1211 Gen\`eve 23, Switzerland.}CERN, 
Gen\`eve, Switzerland.}\\ 

\end{center}

\vspace{2em}


\begin{quotation}
\centerline{\bf ABSTRACT}

\vspace{1em}

We classify possible ``self-duality'' equations for $p$\/-form
gauge fields in space-time dimension up to $D=~16$\/, generalizing the
pioneering work of Corrigan \etal\/~(1982) on Yang-Mills fields ($p=1$\/)
in $4 < D \leq 8$\/. We impose two crucial requirements. First, there
should exist a $2(p+1)$\/-form $T$\/ invariant under a sub-group $H$\/ of
\Sod\/. Second, the \repr\/ for the \Sod\/ curvature of the gauge field
must decompose under $H$\/ in a relevant way. When these criteria are
fulfilled, the ``self-duality'' equations can be candidates as
gauge functions for \Sod\/-covariant and $H$-invariant topological quantum
field theories. Intriguing possibilities occur for $D\geq 10$\/ for various
$p$\/-form gauge fields.
\end{quotation} \end{titlepage} \newpage

\newpage\null


\section{Introduction}

A large class of topological quantum field theories (denoted as $TQFT$\/s)
for a $p$\/-form gauge field~$A_{(p)}$\/ with $(p+1)$\/-form
curvature~$F_{(p+1)}$\/ can be understood as theories which explore the
moduli space of (anti-)\sdes\/
\bea
\label{generic}
\pm F_{\m_1\cdots\m_{p+1}}=\frac{1}{(D/2)!}
{\e_{\m_1\cdots\m_{p+1}}}^{\m_{p+2}\cdots\m_D} F_{\m_{p+2}\cdots\m_D}.
\eea
Assuming that the gauge symmetry of the topological theory is
arbitrary, local redefinitions of the classical fields, the
\sdes\/~(\ref{generic}) are
relevant as topological gauge-fixing functions. With 
the method of $BRST$ quantization, one determines the
$BRST$-invariant Lagrangian to be used in the path integral formulation of
the $TQFT$\/ as well as the observables~\cite{wittentopo},~\cite{bsym}.
Moreover, the topological $BRST$ symmetry can often be (un)twisted
to recover an ordinary \st\/ supersymmetry \cite{wittentwist}, \cite{bks}.

The \sdes\/~(\ref{generic}) rely on the existence of the totally
antisymmetric tensor $\e_{\m_1\cdots \m_D}$\/. Thus, one expects that
Yang-Mills-like $TQFT$\/'s  
can only be constructed for $p$-form gauge fields satisfying the
condition~$2(p+1)=D$\/. However, relaxing some constraints, one foresees
the possibility for new
classes of $TQFT$\/'s involving $p$\/-forms in various
dimensions $D$\/. This seems an interesting perspective to us: it
could provide effective field-theories on the world-volume of branes
(see for instance \cite{bks}, \cite{acha}, \cite{hull}, \cite{duff}); or a
$12$\/-dimensional $TQFT$\/ for a $3$\/-form gauge field, which could be
candidate for the elusive ``$F$\/-theory''.
Furthermore, new instanton solutions could be of interest, as for $p=1$,
$D=8$~\cite{inst}.

Such theories reduce the $D$\/-dimensional Euclidean
{\it invariance} down to invariance under a sub-group $H\subset$~\Sod\/
(with Lie algebra~$h$\/), while preserving the \Sod\/ {\it covariance}.
From a geometrical point of view, in curved space, $H$ would be the
holonomy group of the $D$\/-dimensional manifold over which the
supersymmetric $TQFT$ is constructed. 
This generalization is done by changing
$\e_{\m_1\cdots \m_D}$\/ in (\ref{generic}) into a more general
$2(p+1)$\/-tensor $T$\/. For $D > 2(p+1)$\/, $T$\/ cannot be
invariant (\ie\/
covariantly constant in curved space) under $SO_D$, but rather under one of
its 
sub-groups~$H \subset SO_D$\/. The determination of such invariant tensors 
$T$\/ was first worked out by Corrigan {\it
et al.}\/ in \cite{cfn}, for Yang-Mills fields in dimensions up to
eight.

In the same spirit of~\cite{bks}, our purpose here is thus to exhibit
equations of the type
\bea
\label{f=tf} 
\l F_{\m_1\cdots\m_{p+1}}&=&{T_{\m_1\cdots\m_{p+1}}}^{\n_{1}\cdots\n_{p+1}}
F_{\n_1\cdots\n_{p+1}} \eea where $\l$\/ is to be understood as one of
$T$\/'s eigenvalues. Keeping in mind that we wish (\ref{f=tf}) to be
relevant as a gauge-fixing function, we expect that the field equations
remain a consequence of the Bianchi identity together with
(\ref{f=tf}). This implies that we ought to restrict ourselves to totally
antisymmetric tensors $T$\/. Therefore, equation~(\ref{f=tf}) can be called
a ``generalized self-duality equation'' by analogy with the generic
case~(\ref{generic}) where~$2(p+1)=D$\/. Furthermore, since it only
involves one gauge field, we call~(\ref{f=tf}) of ``pure'' type,
as opposed to self-duality equations that would mix the curvatures for
various gauge fields. 

If a $D$-dimensional $TQFT$ can be constructed from~(\ref{f=tf}), its
Lagrangian will only be $SO_D$\/-{\it covariant}\/ and $H$\/-{\it
invariant}\/.
Its supersymmetry will be a topological $BRST$\/-symmetry, where all the
fermions are ghosts.
However, it may happen that the full $SO_D$\/-invariance be
recovered from a twist of the various fields involved in the theory, and
the ordinary supersymmetry from untwisting the topological $BRST$ symmetry
(a further dimensional reduction can be necessary). An example of this
scenario 
is the one in~\cite{bks}, where a covariant eight-dimensional
$TQFT$\/ is built for a Yang-Mills field, with the $SO_8$\/ 
invariance broken down to a $G_2\subset SO_7\subset SO_8$\/ invariance:
the action explicitly depends on an $SO_8$\/ $4$\/-form $T$, invariant
under $G_2$ only. Nevertheless, untwisting the corresponding Lagrangian
enables to absorb this dependence into field redefinitions, giving rise to
the ordinary Lorentz invariant supersymmetric action in eight dimensions.
The latter is the one that is obtained by dimensional reduction of the
usual $N=1$\/, $D=10$\/ super-Yang-Mills theory~\cite{bks}, \cite{acha}.

Determining $H$\/-invariant tensors $T$\/ can be done using the branching
rules for irreducible representations of semi-simple Lie algebras (we used
the tables in the book from W.G.McKay, J.Patera~\cite{patera}: they provide
the information as to semi-simple algebras, while abelian
factors can be found in~\cite{slansky}).  Furthermore, we are interested in
a situation where a TQFT can be constructed, with the self duality
equation~(\ref{f=tf}) as gauge function. In this perspective, (\ref{f=tf})
must ideally give a number of independent equations on the curvature,
matching  the number $\NN$\/ of gauge invariant components of~$A_{(p)}$\/.
As we shall see, the realization of the requirements towards a $TQFT$\/ is
non trivial, as soon as it comes to cases other than the generic
$2(p+1)=D$\/ for even $D$\/. It even came as a surprise to us that we found
possibilities for other $(p$\/, $D)$\/.

Our paper has two aspects.  On one hand, we give the methodology to probe
the possible relevance of such theories, enlightening the selection of
$h$\/ and $T$\/ for given $(p$\/, $D)$\/'s. On another hand, we implement
the method and explicitly determine the eligible \sas\/~$h$\/ for each
possible value of $p$\/ and $D$\/, using the tables in~\cite{patera}\/
and unpublished results concerning higher dimensional
cases.\footnote{We gratefully thank J. Patera for private communication of
those.}

There are limitations in our analysis. First, the existing tables give
the decompositions of the \Sod\/-representations into maximal semi-simple
sub-algebras. We have thus restricted our analysis to maximal sub-algebras
$h$ (except for $su_{D/2} \subset so_D$ cases that we obtain
explicitly by complexification). 
This could lead us to miss possible
solutions. For instance, the interesting octonionic solution of non-maximal
$g_2\subset so_7\subset so_8$\/ for $(p=1$, $D=8)$ can be extracted from
the solutions 
of the non-maximal (and thus not recorded in~\cite{patera}) sub-algebra
$su_4\subset so_8$. However, $g_2$\/ is maximal within $so_7$\/, and the
octonionic 
solution can nevertheless be found from~\cite{patera}, through a slightly
different and more refined analysis (see subsection~\ref{h}).

Second, we shall not consider in this paper the possible mixing of
curvatures in the self-duality equation~(\ref{f=tf}). Mixed self-duality
equations involving several gauge fields occur, for instance, from
dimensional reduction of a ``pure'' case like~(\ref{f=tf}). An example was
observed in~\cite{bks}: the 
``pure'' $(p=1$, $D=8)$\/ equation gives, when reduced to 4 dimensions, a
non abelian version of the Seiberg-Witten equations, involving a spinor and
a Yang-Mills field. We emphasize that other mixed self-duality equations
may exist (that do not come from dimensional reduction of a ``pure''
case), but we shall not consider them here.

Despite these points, and despite the fact that we only exhibit the
possible $h$\/'s without any certainty that the corresponding $TQFT$\/ can
be given a sense, we nevertheless believe that the results presented
here are already quite interesting by themselves. In particular, they point
out the diversity of the possibilities, depending on the various values of
$D$. They also indicate the specificity of $D=8$ which is the only case
where an exceptional group, $G_2$, arises in our analysis (the only
orthogonal groups that have $G_2$ as a maximal sub-group are $SO_7$\/ and
$SO_{14}$\/, but in the later case, the requirements we impose cannot be
realized). 

The organization is as follows: in section~\ref{groupth} are given the
arguments for the selection of the invariance algebra $h$\/ and for the
conditions on~$T$\/ together with its~$\l$\/'s.  Section~\ref{table}
contains our results: we display the possible invariance \sa\/~$h$\/ for
each pair $(p$\/, $D)$\/ and comment on the peculiarities of each case we
have explored.


\section{Analysis of the Generalized Self-Duality Equation}
\label{groupth}

Our aim for this section is to probe the various criteria to be fulfilled
for the self-duality equation~(\ref{f=tf}) to count for $\NN$\/ gauge
fixing conditions, and to be appropriate for building 
$TQFT$\/'s of various $(p$, $D)$\/.
Let us first consider equation~(\ref{f=tf}) from a general point of view,
keeping in mind that we seek a systematic check of the following
requirements:\\ {\it i)}\/ the \Sod\/ form $T$\/ should be an $H \subset
SO_D$\/-invariant;\\ {\it ii)}\/ the $SO_D$ irreducible representation for
$F$\/ should contain an $\NN$\/-dimensional representation in its
decomposition under $H$;\\ {\it iii)}\/ the eigenvalues $\l$\/ 
should have a suitable degeneracy, allowing~(\ref{f=tf}) to count for
$\NN$\/ independent equations on~$F$\/; \\ {\it iv)}\/ the properties of
$T$\/ should enable to put the action under a Yang-Mills-like form.

We shall detail the points {\it i), ii)}\/ in the following subsection.
Criteria {\it iii)} will be the object of subsection~\ref{iii}, and the 
remaining criteria {\it iv)} of subsection~\ref{iv}.


\subsection{On the selection of the invariance \sa\/~$h$\/}


\label{h}

Before describing the method, let us set up our notations: we consider a
$p$\/-form gauge field in $D$\/ \st\/ dimensions.  Its curvature is a
$(p_F=p+1)$\/-form, and has $d_F={D\choose p_F}$\/ independent
components.  Consequently, the invariant form~$T$\/, relevant to
write~(\ref{f=tf}), is a $p_T$\/-form with $p_T = 2 p_F=2(p+1)$\/.
Since it is chosen totally antisymmetric, $T$ has $d_T=
{D\choose p_T}$\/ independent components. Thus, for a given
$p$\/, we seek a $2(p+1)$\/-form $T$\/ that is invariant under a sub-group
$H$\/ of \Sod\/, while $T$\/ (resp.~$F$\/) lies in a \repr\/ of \Sod\/ of
dimension $d_T$\/ (resp.~$d_F$\/).

The first check {\it i)}\/ is to examine whether an \Sod\/ $p_T$\/-form
can be left 
invariant by the transformations of some \sg\/~$H \subset SO_D$\/. It
consists in inspecting, in~\cite{patera}, the decomposition of the
\sod\/-\repr\/ \trep\/ upon \irr\/ \reprs\/ of all the possible
\sas\/~$h$\/. We then select the $h$\/'s that leave the
$p_T$\/-form invariant, \ie\/ the sub-algebra under which the decomposition
of \trep\/ produces a singlet \rep{1}. Notice that we 
restrict ourselves to maximal sub-algebras, except for the particular
case of a sub-algebra $h$\/ leaving \trep\/ irreducible. In this case, we
pick up the sub-algebra $\tilde{h} \subset h$\/ that leaves $T$\/
invariant. This is the situation of the $(p=1$, $D=8)$\/ case, where the
4-form $T$ can be decomposed into self and anti-self-dual parts,
irreducible under~$so_7$. 
More generally, this occurs for $D=4(p+1)$\/. 

Once such a sub-algebra $h$\/ (or $\tilde{h}$\/) has been selected, we
impose that it satisfies criteria {\it ii)}\/: to end up with a proper
$TQFT$\/ for the $p$\/-form gauge field, the \sdes\/~(\ref{f=tf}) must
count for as many equations as there are of gauge-invariant components
of~$A_{(p)}$\/ (the longitudinal gauge degrees of freedom can be fixed with
the conventional $BRST$\/ formalism for forms~\cite{bsym}). A $p$\/-form
gauge field has a total of $D\choose p$\/ components, but is defined up to
an anticommuting gauge symmetry parameter $\xi_{(p-1)}$\/. The latter has a
total of $D\choose {p-1}$\/ components, but is also defined up to a
commuting gauge symmetry parameter $\xi_{(p-2)}$\/ with $D\choose {p-2}$\/
components, and so on.  Thus, the number $\NN$\/ of gauge-invariant
components of $A_{(p)}$\/ is the number of components of a
$p$\/-form in $(D-1)$\/ dimensions, that is
\bea
\NN = {D\choose {p-1}} -{D\choose {p-2}} + \cdots \pm {D\choose 0}
={{D-1}\choose p}. 
\eea
Therefore, criteria {\it ii)} consists in checking whether the
\sod\/-\repr\/~\frep\/ contains, in its $h$\/-decomposition
\frep\/~=~$\bigoplus$ \rep{d_F}\/$^i$\/ ($i=1 \cdots n$), a \repr\/
\rep{\NN}\/~=~$\bigoplus$ \rep{d_F}\/$^j$ with total dimension $\NN$\/. The
right number of constraints on the gauge field~$A_{(p)}$\/ can then be
obtained by imposing (\ref{f=tf}) if the latter allows to project $F$ on
the space corresponding to \rep{(d_F - \NN)}\/.
In this final sorting of the sub-algebras, we do not impose that
\rep{\NN}\/, nor \rep{(d_F - \NN)}\/, be \irr\/ representations of $h$\/.
Allowing these \reprs\/ to be a sum of irreducible \reprs\/ does not lead
to inconsistencies: (\ref{f=tf}) is then expressed as independant sets of
equations that involve independant combinations of $F$\/'s components.

$T$ can be understood as a linear operator, mapping the space of
$(p+1)$-forms into itself; and since it has been chosen a singlet of $h$\/,
Schur's lemma implies that $T$ acts block diagonally in each of the
\frep\/$^i$\/'s ($T$ being an invariant of $H$\/, its action on an \irr\/
\repr\/ of $H$\/ remains in the same \repr\/). Each
\frep\/$^i$\/ is associated with an eigenvalue $\l^i$, and one can always
choose an appropriate basis to diagonalize $T$ in each
\frep\/$^i$\/ so that it reads:
\bea
\label{diag}
T_{diag} = \left(
\begin{array}{ccccc} [\l^1 {\Bbb{I}}_{(1)}]\\ & [\l^2 {\Bbb{I}}_{(2)}]\\ &&
[\ddots]\\ &&& [\l^i {\Bbb{I}}_{(i)}]\\ &&&& [\ddots] \end{array} \right)
\eea
where ${\Bbb{I}}_{(i)}$ denotes the relevant $({d_F}^i)$\/-dimensional
identity operator involved in the associated sub-space.\footnote{This
actually holds as 
long as each \frep\/$^i$\/ appears only once in the decomposition
\frep\/~=~$\bigoplus$ \rep{d_F}\/$^i$. If a given
\repr\/ has multiplicity $m$ higher than one, then $T$'s action on these
$m$ identical \reprs\/ is one single block (instead of $m$), that is not
necessarily diagonal.}

Taking into account the possible degeneracy of the eigenvalues, the $(\l^1,
\cdots, \l^i, \cdots, \l^n)$\/ take $m$ distinct values $(\m^1, \cdots,
\m^a, \cdots, \m^m)_{(m \leq n)}$\/.  One can then re-write $T$ in terms of
the $\m$\/'s, and define the $h$\/-invariant operator $P_{\m^a}$,
key to the gauge fixing, as $P_{\m^a} = T - \m^a {\Bbb{I}}_{(a)}$.
According to the degeneracy of the $\l$\/'s, each $\m^a$\/ is now
associated to one or a sum of representations (with the same
$\l^j$\/: \rep{d_a}\/ = $\bigoplus$ \rep{d_{F}}$^j$\/), of total dimension 
$d_a$\/.
Imposing $P_{\m^a} F = 0$\/ projects $F$\/ on the sub-space corresponding
to~\rep{d_a}\/.
It amounts to cancelling the projection of $F$ onto all the 
other sub-spaces corresponding to  $\m^b \neq \m^a$\/.
This gives a number of independent equations on $F$\/ equal to the
total dimension $\sum d_b = (d_F - d_a)$\/ of the $(\bigoplus$ \rep{d_b}\/)
representation.  Hence, $P_{\m^a} F =
0$\/ may be suitable as a gauge fixing function in the sense described
above, provided that it counts for $\NN$\/ equations, \ie\/ provided that
$(d_F - d_a) = \NN$\/.

Before taking this any further (we shall come back on this point in
subsection~\ref{iv}), let us first probe the conditions that one has on
the $\l^i$\/'s, and in particular the issue of criteria~{\it iii)}.


\subsection{On the degeneracy of the eigenvalues $\l$}
\label{iii}


Suppose that one has found a sub-algebra $h$\/ and a form $T$\/ satisfying 
criteria {\it i), ii)}\/; and consider only one of the sub-spaces upon
which $F$\/ decomposes. To compute the corresponding eigenvalue $\l^i$\/, 
one can apply $T$\/ on~(\ref{f=tf}) once more to get:
\bea
\label{deuxfoist} 
(\l^i)^2 F_{\m_1\cdots\m_{p+1}}=
{T_{\m_1\cdots\m_{p+1}}}^{\n_1\cdots\n_{p+1}}
{T_{\n_1\cdots\n_{p+1}}}^{\s_1\cdots\s_{p+1}} F_{\s_1\cdots\s_{p+1}}.
\eea
One obtains an equation for $\l^i$\/ if one knows how
${(T^2)_{\m_1\cdots\m_{p+1}}}^{\s_1\cdots\s_{p+1}}$\/ expands upon
$h$\/-invariant tensors, and how these act on $F$\/. Unlike $T$\/, $T^2$\/
lies in a reducible \repr\/ \ttrep\/ of \sod\/: it transforms like the
product of two $p_F$\/-forms, \ttrep\/$\equiv$
\frep\/ $\otimes$ \frep\/, that one first expands as a sum of \irr\/
\reprs\/ of \sod\/.  Second, $T^2$\/ is a tensor of \sod\/, necessarily
$h$\/-invariant because $T$\/ is, so it can only expand as a linear
combination of $h$\/-invariant tensors. One is sure to find, in $T^2$\/, at
least the trace part ${\d_{[\m_1}}^{[\s_1} \cdots
{\d_{\m_{p+1}]}}^{\s_{p+1}]}$\/, as well as a
${T_{\m_1\cdots\m_{p+1}}}^{\s_1\cdots\s_{p+1}}$\/ term, but there might be
others. Generically, let us write the decomposition of \ttrep\/ upon
$h$\/-invariant \irr\/ \reprs\/ of \sod\/ as \bea
\label{t2dec}
\mbox{\ttrep\/} \equiv \mbox{\frep\/$\otimes$\/\frep\/ = \rep{1}\/}
\oplus  \mbox{\trep\/}\oplus \left(\bigoplus_m \mbox{\/\ttrepm}\right).
\eea
Whenever $T$\/ and the trace term are the only tensors to appear
in~(\ref{t2dec}), one has
\bea
\label{ideal}
{(T^2)_{\m_1\cdots\m_{p+1}}}^{\n_1\cdots\n_{p+1}}
= A {\d_{[\m_1}}^{[\n_1} \cdots {\d_{\m_{p+1}]}}^{\n_{p+1}]}
+ B {T_{\m_1\cdots\m_{p+1}}}^{\n_1\cdots\n_{p+1}}
\eea
and the $\l^i$\/'s  are solutions~of
\bea
\label{lambdaequ}
(\l^i)^2 = A(p+1)! + B \l^i,
\eea
where the trace part appears with a non-vanishing coefficient
$A$\/.\footnote{Indeed, if one contracts the remaining indices to get 
${(T^2)_{\m_1\cdots\m_{p+1}}}^{\m_1\cdots\m_{p+1}}$\/, the only term that 
survives is the trace term. 
Furthermore, and since we are considering Euclidean spaces, 
its coefficient $A$\/ cannot be zero unless $T$\/ itself is zero.}
The statement that $A \neq 0$ ensures that $\l^i\neq 0$. 
Equation~(\ref{lambdaequ}) tells us a lot more: as long as
the hypothesis of the unicity of $T$\/ in~(\ref{t2dec}) holds, the
$\l^i$\/'s  can only take two possible values, whatever number of
$h$\/-\irr\/ \reprs\/ is involved in the decomposition of~$F$\/. One can
thus expect some quite restrictive degeneracy of the $\l$\/'s. 

The only possibility for the $\l$\/'s to be allowed more than two values is 
when other invariant tensors than $T$ appear in~(\ref{ideal}).
To each of those invariant tensors $T_m$\/ is associated a set of
eigenvalues. Reproducing the same string of arguments we have used for
$T^2$, one gets the decomposition of each $(T_m)^2$\/ upon the set of
invariant tensors. These lead to a set of equations for the eigenvalues of
$T$ together with the eigenvalues of the $T_m$\/'s.

In general, we expect a high level of degeneracy for the $\l$\/'s.
Unfortunately, to be sure that~(\ref{f=tf}) eventually counts for $\NN$\/
constraints on $F$\/, one would need to know, case by case, the detailed
spectrum of the $\l^i$\/'s. Since the explicit
construction of the $T$ tensor is not our goal here, we shall not take this
any further, but rather turn to requirement~{\it iv)}\/.


\subsection{On the relevance of the self-duality equation towards a $TQFT$}
\label{iv} 


As to the relevance of the \sde\/, there are two things to be kept in mind:
in all the cases we expose in the coming section, the check {\it ii)}\/
ensures that one can always write $\NN$\/ $H$-invariant equations.
However, it is not at all systematic that these equations can be imposed 
with the sole constraint~(\ref{f=tf}). Even if it is the case, it is not
guaranted  that the form $T$\/ is 
suitable to recover a
Yang-Mills like action, plus topological terms, when the gauge function is
squared. Notice that one can always
find an $h$\/-diagonal matrix with the spectrum of eigenvalues suitable to 
project the curvature on the relevant sub-spaces.  However, the later does 
not necessarily correspond to an antisymmetric tensor of \Sod\/ and thus
is not relevant in view of~{\it iv)}.

Criteria {\it iv)} involves non trivial properties of $T$\/. The most
favorable case is when $T$\/ is the only tensor to appear in the right hand
side of~(\ref{t2dec}), and when
$P_{\m^a}F=0$ projects on the relevant sub-space.
The $\NN$ equations coming from~(\ref{f=tf}) can then be written as duality
relations between the ``electric'' and ``magnetic'' parts of the curvature:
\bea
F_{D i_1 \cdots i_p} = {c_{i_1 \cdots i_p}}^{j_1 \cdots j_{p+1}}
F_{j_1 \cdots j_{p+1}},\\
c_{i_1 \cdots i_p j_1 \cdots j_{p+1}} \equiv 
T_{D i_1 \cdots i_p j_1 \cdots j_{p+1}},\nn
\eea
with $(i_1, i_2, \cdots, j_1, \cdots)$ being $(D-1)$-dimensional indices.
Therefore, when the gauge function is squared, using~(\ref{ideal}), one
gets the appropriate 
terms $F\wedge\star F$ and $(\star T) \wedge F \wedge F$ in the action for
the $TQFT$\/: 
\bea
\label{gge1}
&&\left|\l F_{\m_1 \cdots \m_{p+1}} + {T_{\m_1 \cdots \m_{p+1}}}^{\n_1
\cdots \n_{p+1}} F_{\n_1 \cdots \n_{p+1}} \right| ^2 \\
\label{action}
&=& \a \left| F_{\m_1 \cdots \m_{p+1}} \right| ^2 + \b \left({T_{\m_1
\cdots \m_{p+1}}}^{\n_1 \cdots \n_{p+1}} F_{\m_1 \cdots \m_{p+1}}
F_{\n_1 \cdots \n_{p+1}} \right)\\
&\mapsto& \a\, (F\wedge\star F) + \b\, (\star T \wedge F \wedge F)\nn
\eea
with the $[(D-2(p+1))\equiv (D-p_T)]$\/-form $(\star T)$ defined as
$\e_{\m_1\cdots\m_D} T^{\m_1\cdots\m_{p_T}} =
(\star T)_{\m_{p_T + 1}\cdots\m_D}$.
The way supersymmetric terms are added to~(\ref{action}), giving a
$BRST$\/-invariant action, is a straightforward generalization of the work
detailed in~\cite{bks}. 

For less ideal cases, the degeneracy of the $\l$\/'s can have drastic
consequences on the relevance of~(\ref{f=tf}) as a gauge function. Very
often, the self-duality equations amount to a number of
independent conditions exceeding $\NN$\/.  These are the most common
cases, for which the $\NN$\/-dimensional sub-space shares its eigenvalue
$\l$\/ with other eigenspaces.  Another situation is when the
$\NN$\/-dimensional sub-space contains different eigenspaces which are
associated with different eigenvalues $(\l^1\cdots\l^k)$\/. Here, one can
tentatively use as gauge function
\bea
\label{gge2}
\left( \prod_{i=1}^k P_{\l^i} \right) F \equiv \left(
\prod_{i=1}^k \left({T_{\m_1\cdots\m_{p+1}}}^{\n_1\cdots\n_{p+1}} - \l^i
{\d_{[\m_1}}^{[\n_1} \cdots {\d_{\m_{p+1}]}}^{\n_{p+1}]} \right) \right)
F_{\n_1\cdots\n_{p+1}}.
\eea
Whatever the degeneracy of the $\l$\/'s may be, \ie\/ with
either~(\ref{gge1}) or (\ref{gge2}) as gauge function, if the
expansion for $T^2$\/ involves 
other $H$\/-invariant tensors $T_m$\/, the resulting $TQFT$\/ shall have an
unusual Lagrangian, that possibly depends on these other invariant
tensors. Notice that the additional $T_m$\/-terms are not topological,
since the $T_m$\/'s 
do not correspond to totally antisymmetric tensors of \Sod\/; these terms
can nevertheless be of interest.

For a given sub-group $H$ satisfying our criteria, we often found only one
singlet in the decomposition of \trep\/. 
To the later corresponds a given spectrum of the
eigenvalues~$\l^i$\/ (\ie\/ a given set of $\m^a$), often such that the
number of equations obtained on the components of $F$\/ is greater than
$\NN$\/. An example is the case of $h\equiv so_{D-1}$\/, with 
$2(p+1)=(D-1)$\/, in odd space-time dimension~$D$\/. Here,
(\ref{f=tf}) amounts to canceling $F_{D i_1 \cdots i_p}$ ($\NN$\/
equations) {\it and}\/ the $\demi {{D-1}\choose {p+1}}$ anti-self-dual
components of $F_{i_1\cdots i_{p+1}}$\/ ($i_1, i_2, \cdots$ are $(D-1)$\/
dimensional indices). Another example is the $h\equiv so_{d} \times
so_{D-d} \subset so_{D}$ case. Here, the only possible $T$\/ again has
eigenvalues $\pm 1$ and leads to more than $\NN$\/ constraints on the
curvature.

However, there are cases where \trep\/ gives several singlets in its
decomposition under $h$, giving more flexibility as to the spectrum of the
eigenvalues.  The general
solution for $T$\/ is then expressed as a linear combination of the
corresponding tensors: $T=\sum _k \a_k T^{(k)}$\/, with $\l=\sum _k \a_k
\l^{(k)}$\/. Thus, one has the opportunity of adjusting the coefficients
$\a_k$, such that the eventual spectrum allows to project $F$\/ precisely
on the relevant sub-spaces.  The situation of having several $T$'s occurs
in few and privileged cases, but there are some examples.

Notice that relaxing the condition that $h$ be maximal would increase the 
probability to have several independent invariant tensors $T$. Indeed,
let $h'$ be a sub-algebra of $h$, itself being maximally embedded in
$so_D$. In the chain of decompositions of \trep\/, singlet pieces can come
from the various ``upper level'' components of $T$\/: under $h \subset
so_D$\/, \trep\/ decomposes into \rep{1}\/+\/\rep{d_T}\/', and under
$h' \subset h$\/, \rep{d_T}\/' can give rise to additional singlet
pieces.  We shall point out such a privileged situations in the next
section, for instance in the eight-dimensional case.

Following the steps described above, we have obtained non trivial solutions, 
that we are now going to expose.

\newpage
\section{Results and Comments}
\label{table}

We present our various results under the form of a table
exploring the possibilities for $p_T=2(p+1) \leq D$\/ up to $D=16$\/.
This table 
gathers the invariance \sas\/ $h$\/ which fulfill our conditions {\it i),
ii)}.
Each possibly relevant case will be separately detailed afterwards (though,
again, $T$\/ is not explicitly constructed here, so that we are not in a
position to check {\it iv)}\/. 
\vspace{1em}

\noindent
\begin{tabular}{|c||c|c|c|c|c|c|c|}
\hline
&$p=1$&$p=2$&$p=3$&$p=4$&$p=5$&$p=6$&$p=7$\\ 
\hline \hline
$D=4$&\au\/$\,\times\,$\au\/&&&&&& \\ \hline
$D=5$&\au\/$\,\times\,$\au\/&&&&&& \\ \hline
$D=6$&&\at\/&&&&& \\ \hline
$D=7$&\au\/$\,\times\,$\au\/$\,\times\,$\au\/&\at\/&&&&& \\ \hline
$D=8$&$g_2\,\subset\,$\bt\/$^{\star}$&&\dq\/&&&& \\ 
&\at\/&&&&&& \\ \hline
$D=9$&&\au\/$\,\times\,$\at\/&\dq\/&&&& \\ \hline
$D=10$&&\au\/$\,\times\,$\au\/$\,\times\,$\at\/&\aq\/&\dc\/&&& 
\\\hline
$D=11$&\au\/&&\au\/&\dc\/&&& \\ \hline
$D=12$&&\at\/$\,\times\,$\cd\/$\,\subset\,$\bc\/$^{\star}$&\ac\/&&\ds\/
&&\\ 
&&&\au\/$\,\times\,$\ct\/&&&&\\\hline
$D=13$&&&\au\/&&\ds\/ &&\\ \hline
$D=14$&&&&&\as\/&\dse\/ &\\ \hline
$D=15$&\au\/&&\au\/&\at\/&\au\/&\dse\/&\\ 
&&&\at\/&&\at\/&&\\ 
&&&\au\/$\,\times\,$\cd\/&&&&\\ \hline
$D=16$&&&\cd\/$\,\times\,$\cd\/&&\ase\/&&\dh\/\\
&&&\au\/$\,\subset\,$\bse\/$^{\star}$&&\cd\/$\,\times\,$\cd\/&& \\ 
&&&\at\/$\,\subset\,$\bse\/$^{\star}$&&\au\/$\,\times\,$\cq\/&&
\\ 
&&&\au\/$\,\times\,$\cd\/$\,\subset\,$\bse\/$^{\star}$&&&& \\
&&&\bt\/$\,\times\,$\dq\/$\,\subset\,$\bse\/$^{\star}$&&&& \\
\hline
\end{tabular}

{\footnotesize \noindent
The index $(^{\star})$ indicates the particular cases $4(p+1)=D$, where
$T$\/ can be decomposed into self- and anti-self dual parts, so that
\trep\/ is not irreducible under \sod\/, but rather under $so_{(D-1)}$.\\
Let us remind the isomorphisms between simple Lie algebras:
$su_2 \sim so_3 \sim sp_2$\/, $sp_4 \sim so_5$\/, $su_4 \sim
so_6$\/.} 

In the above table, the generic self-duality~(\ref{generic})
systematically appears on the diagonal for even $D=2(p+1)$\/.
Similarly, one has the odd dimensional $(D-1)=2(p+1)$\/ briefly discussed
in sub-section~\ref{iv}.

In $5$, $7$ and $9$ dimensions, the possibilities are of the aforementioned
type $h\equiv so_{D-1}$ or $so_{d} \times so_{D-d}$, for which $\l =
\pm1$\/ and $T$\/ does not project $F$\/ properly: one gets more than
$\NN$\/ constraints on the gauge field. 
Details for the solutions $p=1$\/ of $D=5$, $7$ can be found
in~\cite{cfn}. The other cases with $h\equiv so_{d} \times so_{D-d} $ or
$so_{D-1}$ occurring in higher dimensions are analogous. Therefore, we only
comment on the cases $(D=8$\/, $p=1)$ and $D \geq 10$\/ where more peculiar
sub-algebras $h$\/ occur. Furthermore, we display all the
$h$\/-decompositions of \frep\/~representations in the tables of the
appendix. 

\subsection{$8$-dimensional cases}


Apart from the standard solution (\ref{generic}) of a $3$\/-form, one
can have \at\/ or $g_2$ invariance for the 4-form $T$\/ of the Yang-Mills
case ($p=1$).  These 
are the cases of interest in~\cite{bks}, \cite{acha},
\cite{hull}. 
For $p=1$\/, $D=8$\/, $T$ 
is a $D/2$\/-form and can thus be naturally decomposed into self and
anti-self-dual parts. While \frep\/ = \rep{28}\/, one seeks $\NN = 7$
equations, where the $4$-form $T$, being itself a solution
for~(\ref{generic}), transforms as a \rep{70}\/ $\sim$\/ \rep{35}$^+$\/ +\/
\rep{35}$^-$\/ of $so_8$\/.

We first consider the $g_2 \subset$\/\bt\/ theory.  Under \bt\/,
\rep{35}$^+$\/ and \rep{{35}}$^-$\/ remain irreducible, whereas \frep\/
obviously decomposes into \bea \mbox{\rep{28} = \rep{7} + \rep{21}}.  \eea
Checking the decomposition of \bt\/'s \rep{35}\/ under its sub-algebras
selects $g_2$, under which
\bea
&&\mbox{\rep{35} = \rep{1} + \rep{7} + \rep{27}}\nn \\
&&\mbox{\rep{21} = \rep{7} + \rep{14}}\nn \\
&&\mbox{\rep{7} = \rep{7} : irreducible}.
\eea
The spectrum for $T$\/'s
eigenvalues allows to impose exactly $7$ equations (\ie\/ to project $F$
onto a \rep{7}\/~+~\rep{14}\/).  Indeed, one can check that $T$ is the only
invariant tensor upon which $T^2$ can decompose, so that it admits two
eigenvalues (see sub-section~\ref{iii}), namely $\l_1 = -1$, $\l_2 = 3$
(see~\cite{cfn}). Moreover, $T_{8 i j k}$\/
($i,j,k$\/ are 7-dimensional indices) is antisymmetric and
thus traceless with respect to the $SO_7$ metric. Therefore, the only
possibility for the degeneracy of the $\l$\/'s is the appropriate one:
$21\l_1 + 7 \l_2 = 0$, and imposing~(\ref{f=tf}) with $\l = \l_1$ leads to
the $7$ equations on $F$.  This case was first studied by Corrigan {\it et
al.}\/~\cite{cfn} and the corresponding $TQFT$ was constructed
in~\cite{bks}.  The set of 7 self-duality equations obtained are
\bea
F_{8 i}&=&c_{ijk}F_{jk},\\
c_{ijk}&=&T_{8ijk}\nn
\eea
where the $c_{ijk}$ ($i,j,k = 1\cdots 7$\/) are the structure constants for
octonions. More explicitly, in terms of 8-dimensional components of the
curvature, they read 
\bea
\label{g2}
F_{12} + F_{34} + F_{56} + F_{78} = 0,\nn\\
F_{13} + F_{42} + F_{57} + F_{86} = 0,\nn\\
F_{14} + F_{23} + F_{76} + F_{85} = 0,\nn\\
F_{15} + F_{62} + F_{73} + F_{48} = 0,\nn\\
F_{16} + F_{25} + F_{38} + F_{47} = 0,\nn\\
F_{17} + F_{82} + F_{35} + F_{64} = 0,\nn\\
F_{18} + F_{27} + F_{63} + F_{54} = 0.
\eea

Let us now turn to the case $h\equiv$\/\at\/ (it is not maximal and thus
not recorded 
in the tables from~\cite{patera}). It enables to complexify
the 8-dimensional indices (with $z_a \equiv x_a + i x_{a+4}$,
$a=1\cdots 4$):
\bea
F_{\m\n} &\rightarrow& (F_{z_a z_b}, \ F_{z_a \bar{z}_b},\ 
F_{z_a \bar{z}_a}),\\
\mbox{\rep{28}}&=&\mbox{\rep{6} + \rep{6} + \rep{15} + \rep{1}}.
\eea
To end up with $7$ equations, the projection of $F$\/ on a (\rep{1}\/ +
\rep{6}\/) has to be cancelled. This is possible because there are
three independant \at\/ invariant $4$-forms (see appendix B in~\cite{cfn}),
and one considers $T$ a linear combinaison of them. The coefficients can
be adjusted at will to end up with the self-duality equations: 
\bea
\label{im}
\mbox{(1 real eqs) \hspace*{4em}}&&\sum_{a=1}^4{F_{z_a \bar{z}_a}}=0 \\
\label{six}
\mbox{(6 real eqs) \hspace*{4em}}&&F_{z_a z_b} + \demi \e_{abcd}
F_{\bar{z}_c \bar{z}_d}=0.
\eea
Moreover, one can observe that equation~(\ref{im}) is the imaginary part of
the complex equation 
\bea
\label{compl}
\sum_{a=1}^4 {\left( \pa_{z^a} A_{\bar{z}^a} + \demi [A_{z^a},
A_{\bar{z}^a}] \right) }=0.
\eea
The real part of~(\ref{compl}) gives the Landau gauge condition
$\pa_{\m}A_{\m}=0$. 
It is remarkable that the former equations~(\ref{im})--(\ref{six}),
when re-expressed in terms of
$SO_8$ components $F_{\m\n}$\/, give the set of equations (\ref{g2})
obtained through $g_2$.

Notice that if one considers the maximal $su_4 \times u_1$ rather than
\at\/, only one $T$ is possible, and one cannot adjust its 
eigenvalues. Equation (\ref{f=tf}) then leads to the sets of 13, 16 or 27
equations of~\cite{cfn} and criteria {\it ii)} cannot be satisfied.
This illustrates the situation explained in subsection~\ref{iv}, where more
flexibility comes from relaxing the hypothesis that $h$ be maximal.

\subsection{$10$-dimensional cases}


Here, apart from the case of a 2-form (of the aforementioned type $so_{d}
\times so_{D-d} \equiv so_4 \times so_6$\/),we have as a non-standard
solution a $3$\/-form gauge field with $T$ invariant under
\aq\/ (here again, the maximal sub-group is rather $su_5 \times u_1
\subset so_{10}$).
The $4$\/-form curvature is a~\rep{210}\/, the 8-form $T$\/ is
a~\rep{45}\/ and the self-duality equation should count for $84$ conditions
(\dc\/ has two \rep{210}\/, only one of which is the
$4$\/-form we are interested in). This case can be investigated through
complexification of the 10-dimensional indices (with $z_a\equiv x_a + i
x_{a+5}$):
\bea 
F_{\m\n\r\s} &\rightarrow& (F_{z_a z_b z_c
z_d},\ F_{z_a z_b z_c \bar{z}_d}, \ F_{z_a z_b z_c \bar{z}_c},\ F_{z_a z_b
\bar{z}_c \bar{z}_d},\ F_{z_a z_b \bar{z}_c \bar{z}_b},\ F_{z_a z_b
\bar{z}_a \bar{z}_b}),\\
\mbox{\rep{210}$^A$}&=&\mbox{(\rep{5}\/$^+$\/ + \rep{5}\/$^-$\/)
+ (\rep{40}\/$^+$\/ + \rep{40}\/$^-$\/) + (\rep{10}\/$^+$\/
+ \rep{10}\/$^-$\/) + \rep{75}\/ + \rep{24}\/ +  \rep{1}}.
\eea
An $84$-dimensional sum of representations can be exhibited in various
ways.

\subsection{$11$-dimensional cases}


Two interesting possibilities arise, for a Yang-Mills field as well as a 
$3$-form, both involving the so-called principal~\au\/ of \bc\/ as the
invariance sub-algebra (there are many \au\/ sub-algebras in \sod\/, but
the only one that is maximal is the principal \au\/ for odd~$D$). 
Indeed, our first two requirements are fulfilled: the $1$-form's curvature
is  a \rep{55}\/ and the $3$-form's curvature a \rep{330}\/, while  
\bea
\label{55}
&&\mbox{\rep{55} = \rep{19} + \rep{15} + \rep{11} + (\rep{7} + 
\rep{3}\/)$^{\star}$ \hspace*{5em} for $p=1$,} \\
\label{330}
&&\mbox{\rep{330} = \rep{23} + \rep{21} + \rep{19} + 
$3\times$\rep{17} + $3\times$\rep{13} + $2\times$\rep{11} + 
$3\times$\rep{9} + \rep{7}\/ +\rep{1}} \nn\\
&&\mbox{\hspace*{2em} + (\rep{29} + \rep{25} + \rep{21} + 
$2\times$\rep{15} + $3\times$\rep{5}\/)$^{\star}$ \hspace*{2em} for $p=3$.}
\eea
For the 1-form and 3-form, one needs respectively $\NN =10$ and $120$
equations. There are several posibilities for the corresponding sum of
representations, one of which we point out by~$(\cdots)^{\star}$
in~(\ref{55})--(\ref{330}). 

Notice that one can expect more than two eigenvalues for $T$ in both cases
(other \au\/-invariant tensors appear in the decomposition of $T^2$).
However, as to the $p=3$ case, in
view of the number of representations involved in the decomposition for
\frep\/, we do not know whether the spectrum of the eigenvalues can allow
to project exactly on the complementary representation of $\NN$\/
in~(\ref{330}).

\subsection{$12$-dimensional cases}


In addition to the generic case~(\ref{generic}) of a $5$-form, $D=12$
allows $p=2$\/ and 
$3$\/. For $p=2$, the $(D=8$, 
$p=1)$ pattern is reproduced: since it is a $D/2$-form, $T$ can be
decomposed into self
and anti-self-dual parts (\rep{462}\/$^+$ + \rep{462}\/$^-$), which remain  
irreducible under \bc\/.
Moreover, the $3$\/-form curvature lies in the \rep{220}\/ of $so_{12}$\/, 
which, under \bc\/, automatically gives rise to the crucial $\NN=55$ 
dimensional \repr\/:
\bea
\mbox{\rep{220}} &=& \mbox{\rep{55} + \rep{165}}.
\eea
Inspecting \trep\/'s decompositions under \bc\/'s maximal sub-algebras
selects \at\/$\times$\cd\/ $\sim$ \dt\/$\times$\bd\/.
The corresponding \reprs\/ for the curvature $F$ behave under
\at\/$\times$\cd\/ as:
\bea
\mbox{\rep{55}}&=&
\mbox{(\rep{15},\rep{1}) + (\rep{6},\rep{5}) + (\rep{1},\rep{10})}\\
\mbox{\rep{165}}&=&\mbox{(\rep{10}\/$^+$,\rep{1}) +
(\rep{10}\/$^-$,\rep{1}) + (\rep{15},\rep{5}) +
(\rep{6},\rep{10}) +  (\rep{1},\rep{10})}.
\eea
Unlike the $(p=1$, $D=8)$\/ case, $T$\/ is not the only invariant tensors 
to appear in $T^2$\/ so there can be more than two values for $\l$\/. 
We do not know whether their spectrum enables to get the wanted 55 
equations from~(\ref{f=tf}).

The remaining case $p=3$ admits two possible invariance sub-algebras,
namely \ac\/ (non-maximal) and \au\/$\times$\/\ct\/.  The decomposition for
\frep\/~=~\trep\/~=~\rep{495} under \ac\/ is obtained by complexification
of the indices and an $\NN=165$-dimensional representation can be
extracted. As for the case of $p=2$\/, other invariant tensors appear in
the right hand side of~(\ref{t2dec}).

\subsection{$13$- and $14$- dimensional cases}


As for the 11-dimensional 1- and 3-forms, for $p=3$\/ in 13 dimensions, it
is the principal \au\/ that arises.
The representation \rep{1287}\/ for the 8-form $T$ has two singlet
components under \au\/. This suggests that one has some freedom to arrange
the eigenvalues, but we do not know whether 
$F$\/'s projection onto the $\NN=220$-dimensional reducible representation
can be cancelled.

The $(D=14$, $p=5)$\/ \as\/ possibility is another case of
complexification of the indices and is easily worked out: $T$\/ has only
one singlet component and an $\NN = 1287$-dimensional representation can be
found in the decomposition for \frep\/~=~\rep{3003}\/.

\subsection{$15$-dimensional cases}


$D=15$ offers quite a few possibilities, for all possible form degree
except the 2-form gauge field.
The 6-form is of the aforementioned type $h\equiv so_{(D-1)}$\/.

For the case of a Yang-Mills field, $F$ lies in the representation
\rep{105} of \bse\/ and $T$ in the \rep{1365}. The later gives two 
singlets in its decomposition under the maximal \au\/ $\subset$ \bse\/.

In the case of a 3-form gauge field, three maximal sub-algebras are
suitable, namely, \au\/, \at\/ and \au\/$\,\times\,$\cd\/. Here, \trep\/ =
\rep{6435} gives four singlets under \au\/ against only one under the
last two.  The decompositions of \frep\/~= \rep{1365} can be found in
the appendix, where the $\NN=364$-dimensional representation is singled
out.  It involves an increasing number of low-dimensional representations
of $h$\/, thus the fulfilling of {\it iv)} seems more and more difficult to
us.  However, for the case of \au\/, $T$ being a linear combination of four
invariant tensors, there is more freedom for a relevant choice of the
eigenvalues.

\subsection{$16$-dimensional cases}


In 16 dimensions, for $p=3$\/, one has the case of $4(p+1)=D$.
It gives several possible sub-algebras $h\subset so_{15}\subset
so_{16}$\/. 
Unlike the eight and twelve dimensional analogues, here $T$ also contains a
singlet in its decomposition under \cd\/$\times$\cd\/, 
maximal within $so_{16}$. 
Finally, one has a 5-form with either \ase\/ (complexification of the
indices), \cd\/$\times$\cd\/ or \au\/$\times$\cq\/.
For all these cases, $T$\/ has only one singlet component.


\section{Conclusion}


Recent developments have indicated that $TQFT$'s in   various  
dimensions, \ie\/ theories with
twisted supersymmetries,  may  play an important role.   The case of  
the eight dimensional Yang-Mills  $TQFT$
has indicated the relevance of a theory defined on a manifold with
special holonomy group $H\subset SO_D$. The key to such $TQFT$'s is the
existence of $H$\/-invariant 
``self-duality equations''~(\ref{f=tf}) for the curvature of the gauge
field: via $BRST$ quantization, they can serve as gauge fixing functions,
provided they satisfy certain restrictions. 

In this paper, we have pointed out restrictive conditions to be fulfilled 
by~(\ref{f=tf}) to possibly admit instanton-like solutions and determine
$D$-dimensional $TQFT$'s.
We have not restricted ourselves to Yang-Mills fields, but have
considered $p$-form gauge fields, in view of their importance in the
definition of field theories for branes.
Following the first steps exhibited in our analysis, we have
established a table, listing for each $p$ and $D \leq
16$,  the possible subgroups $H\subset SO_D$\/ for which an
\Sod\/-covariant and 
$H$\/-invariant self-duality equation can exist. 
We did not go any further in building the $TQFT$'s, 
but we believe that for some high dimensional cases,
interesting and possibly unfamiliar theories could arise.

\vspace{1em}

\noindent {\large {\bf Acknowledgments:}}

We are grateful to R.~Stora for interesting discussions related to 
this work. We also thank O.~Babelon, E.~Ragoucy, P.~Roche, M.~Talon for
fruitful comments, and J.~Patera for communication of unpublished
results.

\newpage
\centerline{\bf{\sc Appendix}}


\def\xx{\begin{minipage}{46em}\vspace{-1em} \flushleft \bea}
\def\yy{\nn \eea \vspace{-2em} \end{minipage} \\ \hline}
\def\yyy{\nn \eea \vspace{-2em} \end{minipage} \\ \cline{2-2}}
\def\zz{\nn \eea \end{minipage} \\ & \xx}

\noindent
{\small Here, we give the $h$\/-decomposition of \frep\/ for all the
relevant cases for $D\geq 10$\/; one possibility for the $\NN$-dimensional
reducible representation is
singled out with $(\cdots)^{\star}$.}

{\scriptsize
\vspace{1em}

\noindent
\begin{tabular}{|c|c|}
\hline
$ D=10$ & $p=2$, $d_F=120$, $d_T=210$, $\NN=36$\\ 
\hline 
\au\/$\times$\au\/$\times$\at\/&\xx \mbox{\rep{120}=[(\rep{1},\rep{3
},\rep{6})+(\rep{3},\rep{1},\rep{6})]$^{\star}$+
(\rep{2},\rep{2},\rep{15})+(\rep{1},\rep{1},\rep{10}$^+$)+
(\rep{1},\rep{1},\rep{10}$^-$)+(\rep{2},\rep{2},\rep{1})}
\yy
& $p=3$, $d_F=210$, $d_T=45$, $\NN=84$ \\ 
\hline 
\aq\/&\xx \mbox{\rep{210}=(\rep{40}\/$^{\pm}$\/+\rep{10}\/$^+
$\/+\rep{10}\/$^-$\/+\rep{24}\/)$^{\star}
$+\rep{75}\/+\rep{40}\/$^{\mp}$\/+\rep{5}\/$^+$\/+\rep{5}\/$^-
$\/+\rep{1}}
\yy
\hline
$D=11$ & $p=1$, $d_F=55$, $d_T=330$, $\NN=10$ \\ 
\hline 
\au\/&\xx \mbox{\rep{55}=(\rep{7}+
\rep{3}\/)$^{\star}$+\rep{19}+\rep{15}+\rep{11}}
\yy
& $p=3$, $d_F=330$, $d_T=165$, $\NN=120$\\ 
\hline 
\au\/&\xx \mbox{\rep{330}=(\rep{29}+\rep{25}+\rep{21}+
$2\times$\rep{15}+$3\times$\rep{5}\/)$^{\star}$+\rep{23}+\rep{21}+
\rep{19}+$3\times$\rep{17}+$3\times$\rep{13} + $2\times$\rep{11}+
$3\times$\rep{9}+\rep{7}\/+\rep{1}}
\yy
\hline
$D=12$& $p=2$, $d_F=220$, $d_T=462+462$, $\NN=55$\\ 
\hline 
\bse\/&
\xx \mbox{\rep{220} = \rep{55}$^{\star}$+\rep{165}}\yyy
$\supset$\at\/$\times$\cd\/ & \xx \mbox{\rep{55}=
(\rep{15},\rep{1})+(\rep{6},\rep{5})+(\rep{1},\rep{10}),} 
&&\mbox{\rep{165}=(\rep{10}\/$^+$,\rep{1})+
(\rep{10}\/$^-$,\rep{1})+(\rep{15},\rep{5})+
(\rep{6},\rep{10})+ (\rep{1},\rep{10})}
\yy
& $p=3$, $d_F=495$, $d_T=495$, $\NN=165$\\ 
\hline 
\ac\/&\xx \mbox{\rep{495}=(\rep{105}'$^{\pm}$
+2$\times$\rep{15}$^+$+2$\times$\rep{15}$^-$)$^{\star}$
+\rep{189}+\rep{150}'$^{\mp}$+\rep{35}+\rep{1}}
\yy
\au\/$\times$\ct\/&\xx \mbox{\rep{495}=[(\rep{1},\rep{90})+
(\rep{5},\rep{14}')+(\rep{5},\rep{1})]$^{\star}$
+ (\rep{3},\rep{70})+(\rep{3},\rep{21})+
(\rep{3},\rep{14}')+(\rep{1},\rep{14}')+(\rep{1},\rep{1})}
\yy
\hline
$D=13$ & $p=3$, $d_F=715$, $d_T=1287$, $\NN=220$\\ 
\hline 
\au\/&\xx \mbox{\rep{715}$^A$=(\rep{37}+\rep{33}+\rep{31}+2$
\times$\rep{29}+\rep{27}+\rep{25}+\rep{9})$^{\star}$+2$
\times$\rep{25}+2$\times$\rep{23}+4$\times$\rep{21}+3$
\times$\rep{19}+4$\times$\rep{17}}\zz
\mbox{+3$\times$\rep{15}+5$
\times$\rep{13}+2$\times$\rep{11}+3$\times$\rep{9}+2$
\times$\rep{7}+3$\times$\rep{5}+2$\times$\rep{1}}
\yy
\hline
$D=14$ & $p=5$, $d_F=3003$, $d_T=91$, $\NN=1287$\\ 
\hline 
\as\/&\xx\mbox{\rep{3003}=(\rep{490}$_{(1)}^+$+\rep{392}\/+\rep{22
4}$^+$+\rep{112}$^+$+\rep{48}+\rep{21}$^+$\/)$^{\star}$
+\rep{784}+\rep{490}$_{(1)}^-$\/+\rep{22
4}$^-$+\rep{112}$^-$+(\rep{35}$^+$+\rep{35}$^-$\/)+\rep{21}$^
-$+\rep{7}$^+$\/+\rep{7}$^-$+\rep{1}\/}
\yy 
\hline
$D=15$& $p=1$, $d_F=105$, $d_T=1365$, $\NN=14$\\ \hline 
\au\/&\xx \mbox{\rep{105} = (\rep{11}+\rep{3})$^{\star}$
+ \rep{27}+\rep{23}+\rep{19}+\rep{15}+\rep{7}}
\yy
& $p=3$, $d_F=1365$, $d_T=6435$, $\NN=364$\\ 
\hline 
\au\/&\xx \mbox{\rep{1365} = (\rep{45}+\rep{41}+\rep{39}+
2$\times$\rep{37}+\rep{35}+3$\times$\rep{33}+\rep{31} )$^{\star}$
+\rep{31}+4$\times$\rep{29}+3$\times$\rep{27}+5$\times$\rep{25}}\zz
\mbox{+4$\times$\rep{23}+6$\times$\rep{21}+4$\times$\rep{19}+6$
\times$\rep{17}+4$\times$\rep{15}+6$\times$\rep{13}+3$\times
$\rep{11}+5$\times$\rep{9}+2$\times$\rep{7}+4$\times$\rep{5}+2$
\times$\rep{1}}
\yy
\at\/&\xx \mbox{\rep{1365}=(\rep{105}+\rep{175}+
\rep{84})$^{\star}$+(\rep{256}$^+$+\rep{256}$^-$)+\rep{175}+
\rep{84}+(\rep{35}$^+$+\rep{35}$^-$)+(\rep{45}$^+$+
\rep{45}$^-$)+2$\times$\rep{20}+2$\times$\rep{15} }
\yy
\au\/$\times$\cd\/&\xx \mbox{\rep{1365} = [(\rep{5},\rep{35}')+
(\rep{3},\rep{35})+(\rep{5},\rep{14})+
(\rep{1},\rep{14})]$^{\star}$+(\rep{3},\rep{81})+
(\rep{7},\rep{35})+(\rep{5},\rep{35})+
(\rep{1},\rep{35}')}\zz
\mbox{+(\rep{7},\rep{10})+(\rep{9},\rep{5})+(\rep{3},\rep{14})+
(\rep{5},\rep{10})+2$\times$(\rep{3},\rep{10})+(\rep{5},\rep{5})
+ (\rep{1},\rep{5})+(\rep{5},\rep{1})+(\rep{1},\rep{1})}
\yy
\end{tabular}

\noindent
\begin{tabular}{|c|c|}
\hline
&$p=4$, $d_F=3003$, $d_T=3003$, $\NN=1001$\\ 
\hline 
\at\/&\xx \mbox{\rep{3003}=(\rep{300}+\rep{280}$^{\pm}
$+\rep{105}+\rep{175}+\rep{45}$^+$+\rep{45}$^-$+\rep{20}+2$
\times$\rep{15}+\rep{1})$^{\star}$}\zz
\mbox{+\rep{280}$^{\mp}$+2$\times$\rep{256}$^+
$+2$\times$\rep{256}$^-$+2$\times$\rep{175}+2$\times$\rep{84}+2$
\times$\rep{45}$^+$+2$\times$\rep{45}$^-$}
\yy
& $p=5$, $d_F=5005$, $d_T=455$, $\NN=2002$\\ 
\hline 
\au\/&\xx \mbox{\rep{5005}=(\rep{55}+\rep{51}+\rep{49}+2$
\times$\rep{47}+2$\times$\rep{45}+4$\times$\rep{43}+3$
\times$\rep{41}+6$\times$\rep{39}+5$\times$\rep{37}+8$
\times$\rep{35}+8$\times$\rep{33}+12$\times$\rep{31}+\rep{2
3}+\rep{7}+\rep{3})$^{\star}$+\rep{37}+10$\times$\rep{29}}\zz
\mbox{+14$\times$\rep{27}+13$\times$\rep{25}+15$\times$\rep{23}+14$
\times$\rep{21}+18$\times$\rep{19}+13$\times$\rep{17}+17$
\times$\rep{15}+13$\times$\rep{13}+14$\times$\rep{11}+9$
\times$\rep{9}+11$\times$\rep{7}+4$\times$\rep{5}+5$\times$\rep{3}}
\yy
\at\/&\xx \mbox{\rep{5005} = (2$\times$\rep{300}+\rep{256}$^+$+
\rep{256}$^-$+5$\times$\rep{175}+\rep{15})$^{\star}$+
\rep{729}+2$\times$(\rep{280}$^+$+\rep{280}$^-$)}\zz
\mbox{+(\rep{256}$^+$+\rep{256}$^-$)+(\rep{35}$^+$+\rep{35}$^-$) 
+ 3$\times$\rep{84}+3$\times$(\rep{45}$^+$+\rep{45}$^-$)+
\rep{20}+2$\times$\rep{15}}
\yy
\hline
$D=16$ & $p=3$, $d_F=1820$, $d_T=6435+6435$, $\NN=455$\\ 
\hline 
\cd\/$\times$\cd\/&\xx \mbox{\rep{1820} = [(\rep{35},\rep{10})+
(\rep{10},\rep{10})+(\rep{5},\rep{1})]$^{\star}$+(\rep{14},\rep{14})+
(\rep{10},\rep{35})+(\rep{35},\rep{5})+(\rep{5},\rep{35})
+(\rep{14},\rep{5})}\zz
\mbox{+(\rep{5},\rep{14})+(\rep{35}',\rep{1})+(\rep{1},\rep{35}')
+(\rep{10},\rep{10})+(\rep{10},\rep{5})+(\rep{5},\rep{10})+
(\rep{14},\rep{1})+(\rep{5},\rep{5})+(\rep{1},\rep{14})+(\rep{1},\rep{5})+ 
(\rep{1},\rep{1})\hspace{2em}}
\yy
\bse\/&\xx \mbox{\rep{1820} = \rep{1365}+\rep{455}} \yyy
$\supset$\au\/&\xx \mbox{\rep{455}=\rep{37}+\rep{33}+\rep{31}+
\rep{29}+\rep{27}+2$\times$\rep{25}+\rep{23}+2$\times$\rep{21}
+ 2$\times$\rep{19}+2$\times$\rep{17}+2$\times$\rep{15}+3$
\times$\rep{13}+\rep{11}+2$\times$\rep{9}+\rep{7}+
\rep{5}+\rep{1}}\zz
\mbox{\rep{1365}=\rep{45}+\rep{41}+\rep{39}+
2$\times$\rep{37}+\rep{35}+3$\times$\rep{33}+2$\times$\rep{31}
+ 4$\times$\rep{29}+3$\times$\rep{27}+5$\times$\rep{25}+
4$\times$\rep{23}+6$\times$\rep{21}+4$\times$\rep{19}}\zz
\mbox{+6$\times$\rep{17}
+ 4$\times$\rep{15}+6$\times$\rep{13}+3$\times$\rep{11}+
5$\times$\rep{9}+2$\times$\rep{7}+4$\times$\rep{5}+2$\times$\rep{1}}
\yyy
$\supset$\at\/&\xx \mbox{\rep{455}=\rep{175}+\rep{84}
+\rep{35}$^+$+\rep{35}$^-$+\rep{45}$^+$+\rep{45}$^-$+\rep{20}
+\rep{15}+\rep{1}}\zz
\mbox{\rep{1365}=\rep{256}$^+$+\rep{256}$^-$+\rep{105}+
2$\times$\rep{175}+2$\times$\rep{84}+\rep{35}$^+$+\rep{35}$^-$+
\rep{45}$^+$+\rep{45}$^-$+2$\times$\rep{20}+2$\times$\rep{15}}
\yyy
$\supset$\au\/$\times$\cd\/&\xx \mbox{\rep{455}=
(\rep{5},\rep{35})+(\rep{1},\rep{30})+
(\rep{3},\rep{35})+(\rep{7},\rep{10})+(\rep{3},\rep{10})
+(\rep{5},\rep{5})+(\rep{3},\rep{5})+(\rep{1},\rep{5})} \zz
\mbox{\rep{1365}=(\rep{3},\rep{81})+(\rep{5},\rep{35}')
+(\rep{5},\rep{35})+(\rep{7},\rep{35})+(\rep{1},\rep{35}')
+(\rep{3},\rep{35})+(\rep{5},\rep{14})+(\rep{7},\rep{10})}\zz
\mbox{+(\rep{9},\rep{5})+(\rep{3},\rep{1
4})+(\rep{5},\rep{10})
+(\rep{1},\rep{14})+2$\times$(\rep{3},\rep{10})+(\rep{5},\rep{5})
+(\rep{1},\rep{5})+(\rep{5},\rep{1})+(\rep{1},\rep{1})} \yyy
$\supset$\bt\/$\times$\dq\/&\xx \mbox{\rep{455}=(\rep{7},\rep{28})
+(\rep{21},\rep{8}$^V$)+(\rep{1},\rep{56}$^C$)+(\rep{35},\rep{1})}\zz
\mbox{\rep{1365}=(\rep{21},\rep{28})+(\rep{7},\rep{56}$^C$)
+(\rep{35},\rep{8}$^V$)+(\rep{1},\rep{35}$^S$)+(\rep{1},\rep{35}$^C$)
+(\rep{35},\rep{1})}
\yy
& $p=5$, $d_F=8008$, $d_T=1820$, $\NN=3003$\\ 
\hline 
\ase\/&\xx \mbox{\rep{8008}=(\rep{2352}+\rep{420}$^{\pm
}$+\rep{70}+\rep{70}+\rep{63}+\rep{28}$^{\pm}$)$^{\star}
$+(\rep{1512}$^+$+\rep{1512}$^-$)}\zz
\mbox{+\rep{720}+(\rep{378}$^
+$+\rep{378}$^-$)+\rep{420}$^{\mp}$+\rep{28}$^{\mp}$+(\rep{28}$^+
$+\rep{28}$^-$)+\rep{1}}
\yy
\cd\/$\times$\cd\/&\xx \mbox{\rep{8008}=[(\rep{3
5},\rep{35})+(\rep{81},\rep{5})+
(\rep{10},\rep{35}')+(\rep{5},\rep{81})+(\rep{14},\rep{14})
+(\rep{81},\rep{1})+(\rep{1},\rep{81})+(\rep{14},\rep{10})+(\rep{1
0},\rep{10})+2$\times$(\rep{10},\rep{1})]$^{\star}$}\zz
\mbox{+(\rep{35}',\rep{10})+(\rep{30},\rep{10})+(\rep{10},\rep{3
0})+(\rep{35},\rep{14})+(\rep{14},\rep{35})+2$
\times$(\rep{35},\rep{10})+2$
\times$(\rep{10},\rep{35})+2$\times$(\rep{35},\rep{5})+2$
\times$(\rep{5},\rep{35})}\zz
\mbox{+(\rep{14},\rep{10})+2$
\times$(\rep{10},\rep{14})+(\rep{5},\rep{14})+(\rep{1
4},\rep{5})+(\rep{35},\rep{1})+(\rep{1},\rep{35})+3$
\times$(\rep{10},\rep{5})+3$\times$(\rep{5},\rep{10})+
(\rep{5},\rep{5})+2$\times$(\rep{1},\rep{10})}
\yy
\au\/$\times$\cq\/&\xx \mbox{\rep{8008}=[(\rep{5},\rep{315})+
(\rep{3},\rep{315})+(\rep{5},\rep{42})+(\rep{7},\rep{27})+
(\rep{3},\rep{27})+(\rep{3},\rep{1})]$^{\star}$+(\rep{1},\rep{825})
+(\rep{3},\rep{792}$^{(1)}$)}\zz
\mbox{(\rep{3},\rep{308})+(\rep{1},\rep{315})+(\rep{5},\rep{36})
+(\rep{3},\rep{42})+(\rep{5},\rep{27})+(\rep{1},\rep{36})+
(\rep{3},\rep{27})+(\rep{7},\rep{1})}
\yy
\end{tabular}}


\newcommand{\np}[1]{Nucl.\ Phys.~{\bf #1}}
\newcommand{\pl}[1]{Phys.\ Lett.~{\bf #1}}
\newcommand{\cmp}[1]{Comm.\ Math.\ Phys.~{\bf #1}}
\newcommand{\pr}[1]{Phys.\ Rev.~{\bf #1}}
\newcommand{\prl}[1]{Phys.\ Rev.\ lett.~{\bf #1}}
\newcommand{\ptp}[1]{Prog.\ Theor.\ Phys.~{\bf#1}}
\newcommand{\ptps}[1]{Prog.\ Theor.\ Phys.\ suppl.~{\bf #1}}
\newcommand{\mpl}[1]{Mod.\ Phys.\ Lett.~{\bf #1}}
\newcommand{\ijmp}[1]{Int.\ Jour.\ Mod.\ Phys.~{\bf #1}}
\newcommand{\jp}[1]{Jour.\ Phys.~{\bf #1}}
\newcommand{\jmp}[1]{Jour.\ Math.\ Phys.~{\bf #1}}
\def\hh{\hbox{hep-th/}}

\end{document}